\documentclass[%
 reprint,
superscriptaddress,
 amsmath,amssymb,
 aps,
showkeys,
]{revtex4-2}

\usepackage{graphicx}
\usepackage{dcolumn}
\usepackage{bm}
\usepackage[italicdiff]{physics}
\usepackage{amsmath, mathtools, amssymb, ascmac, fancybox, xcolor}
\usepackage[colorlinks=true,linkcolor=teal,citecolor=teal,urlcolor=blue]{hyperref}

\begin{document}

\preprint{APS/123-QED}

\title{Nonlinear Drude weight of the one-dimensional Hubbard model}


\author{Tetsuya Iwasaki}
\affiliation{Department of Physics, Graduate School of Science, The University of Tokyo, Tokyo~113-0033, Japan}

\author{Hosho Katsura}
\affiliation{Department of Physics, Graduate School of Science, The University of Tokyo, Tokyo~113-0033, Japan}
\affiliation{Institute for Physics of Intelligence, The University of Tokyo, Tokyo 113-0033, Japan}
\affiliation{Trans-Scale Quantum Science Institute, The University of Tokyo, Tokyo 113-0033, Japan}

\date{\today}

\begin{abstract}
We investigate nonlinear Drude weights (NLDWs) in the one-dimensional repulsive Hubbard model at zero temperature by combining exact Bethe-ansatz calculations with low-energy effective field theory.
At quarter filling, we first derive the strong-coupling expansion of the NLDWs and confirm it numerically over a wide range of interaction strength.
We then compare the numerical results with the prediction of the Tomonaga-Luttinger liquid (TLL) description including irrelevant perturbations.
While band-curvature corrections yield finite contributions to higher-order NLDWs, the Umklapp interaction predicts divergent NLDWs when the order $n$ of the Drude weight exceeds a threshold determined by the TLL parameter. 
In contrast, finite-size scaling of the exact Bethe-ansatz results indicates that all calculated NLDWs remain finite in the thermodynamic limit, revealing a discrepancy between the exact results and the predictions of the low-energy effective field theory.
At half filling, we analyze the finite-size scaling of the NLDWs across the Mott metal-insulator transition.
We derive their asymptotic behavior in the insulating phase and propose a hyperscaling ansatz for NLDWs near the critical point, which is verified numerically.
Our results clarify the interaction dependence and critical scaling of nonlinear transport coefficients in the one-dimensional Hubbard model and highlight limitations of the conventional low-energy effective description for higher-order transport.
\end{abstract}

\maketitle

\section{\label{sec:intro}Introduction}

Understanding transport in interacting quantum many-body systems is a central problem in condensed matter physics \cite{NagaosaSinovaOnodaMacDonaldOng,ImadaFujimoriTokura,Bertini}.
One-dimensional quantum systems provide an ideal platform for investigating transport
because powerful analytical techniques, including the Bethe ansatz \cite{Takahashi, Essler}, conformal field theory (CFT) \cite{yellow}, and bosonization \cite{Giamarchi} are available.
These methods have revealed a variety of unconventional transport phenomena in strongly correlated quantum systems \cite{Bertini,ZotosNaefPrelovsek}.
Among various transport coefficients,
the Drude weight plays a particularly important role as a measure of ballistic transport \cite{Kohn, Resta}.
It has been extensively studied in one-dimensional quantum systems, particularly in integrable systems \cite{IlievskiDeNardis,ZotosNaefPrelovsek,Bertini,ShastrySutherland, Zotos,PeresSacramentoCampbellCarmelo,LiuYinZhangZhangGuan}.
One of the most extensively studied systems in this context is the one-dimensional Hubbard model.

The one-dimensional Hubbard model describes interacting electrons on a lattice and serves as a paradigmatic model of strongly correlated electron systems \cite{Essler, Deguchietal}.
At half filling (electron density $\nu=1$), the ground state is a Mott insulating phase \cite{LiebWu,ShastrySutherland}, whereas away from half filling the low-energy physics is described by Tomonaga-Luttinger liquid (TLL) theory \cite{Haldane1, Giamarchi, solyom, FrahmKorepin, KawakamiYang}.
The Drude weight of the one-dimensional Hubbard model has been investigated by various analytical approaches \cite{FujimotoKawakami,ShastrySutherland,YuFowler,FrahmKorepin,Schulz,StaffordMillisShastry,StaffordMillis,LuoBasakPuGuan,KirchnerEvertzHanke}.
In the metallic phase, the Drude weight is expressed in terms of the TLL parameters \cite{Giamarchi, FrahmKorepin} and can be determined exactly from the Bethe ansatz solution \cite{Schulz, FujimotoKawakami}.
Furthermore, critical scaling near the Mott transition and finite-size scaling of the Drude weight in the insulating phase have also been elucidated \cite{StaffordMillisShastry, FyeMartinsScalapinoetal, StaffordMillis}.
As a result, the linear-response transport properties of the one-dimensional Hubbard model have been understood in considerable detail.

Despite this remarkable progress, the understanding of transport has been largely confined to the linear-response regime.
Characterizing ballistic transport beyond linear response therefore remains an important problem.
In this context, the nonlinear Drude weights (NLDWs) have been introduced as a natural extension of the conventional Drude weight to characterize ballistic transport beyond linear response \cite{WatanabeOshikawa, WatanabeLiuOshikawa}.
Their properties have been investigated most extensively in the spin-$\frac12$ XXZ chain \cite{WatanabeOshikawa,TanikawaTakasanKatsura,TanikawaKatsura,UrichukKlumperSirker}.
The third-order Drude weight was first obtained analytically from the Bethe ansatz solution \cite{WatanabeOshikawa}.
The same work also reported divergent behavior for certain ranges of the anisotropy parameter.
Subsequent studies extended the analytical calculations to the fifth order Drude weight \cite{TanikawaTakasanKatsura} and further showed that NLDWs divergences at arbitrary orders \cite{TanikawaTakasanKatsura,TanikawaKatsura}.
These divergences have been attributed to the Umklapp interaction in the low-energy effective field theory \cite{TanikawaTakasanKatsura,FukusumiBarisic}.
Similar singular behavior has also been reported in free-electron systems with a single defect \cite{TakasanOshikawaWatanabe},
indicating that divergence of NLDWs can arise in a broader class of one-dimensional quantum systems.
The same work also showed that evaluating the NLDWs directly from the nonlinear conductivity, rather than via the generalized Kohn formula, removes these apparent divergences \cite{TakasanOshikawaWatanabe}.
These studies have established NLDWs as a useful probe of nonlinear transport in integrable quantum systems.
While NLDWs have been investigated in both integrable spin chains and noninteracting electron systems, their properties in interacting electron systems remain largely unexplored.

To fill this gap, we investigate NLDWs at zero temperature in the one-dimensional Hubbard model.
We calculate the NLDWs in the metallic phase of 
this model.
We then compare the numerical results with the predictions of the low-energy effective theory based on bosonization.
While the presence of Umklapp interactions would lead to divergences of NLDWs in the thermodynamic limit, we find no such evidence, suggesting that the corresponding Umklapp processes may be absent at quarter filling.
We further investigate the finite-size scaling of NLDWs near half filling by extending the scaling theory of Stafford and Millis \cite{StaffordMillis} to higher-order responses.
We demonstrate that the finite-size scaling of the NLDWs is given by a power law multiplied by an exponential decay in the Mott insulating phase.
We also identify a hyperscaling form for the NLDWs in the vicinity of the Mott transition.
Our results provide new insight into nonlinear transport in interacting electron systems.

The remainder of this paper is organized as follows.
In Sec.~\ref{sec:model}, we introduce the one-dimensional Hubbard model, briefly review the Bethe ansatz solution, and define the NLDWs.
In Sec.~\ref{sec:quarter}, we first present numerical calculations of NLDWs as functions of the interaction strength through seventh order and then analyze higher-order NLDWs to assess the existence of the corresponding Umklapp processes at quarter filling.
In Sec.~\ref{sec:scaling}, we investigate the finite-size scaling of NLDWs by extending the scaling theory of Stafford and Millis to higher-order responses, and examine their hyperscaling behavior near the Mott transition.
Finally, we summarize our results and discuss future perspectives in Sec.~\ref{sec:discussion}. 
Additional technical details, including the strong-coupling analysis, field-theoretical derivations of the Umklapp contribution, the finite-size scaling analysis at half filling, supplementary numerical results, and a comparison with the XXZ chain at finite magnetization, are presented in the Appendices.

\section{\label{sec:model}Nonlinear Drude weights in the one-dimensional Hubbard model}

\subsection{\label{subsec:model}Model}
We consider the one-dimensional Hubbard model with $L$ sites threaded by a magnetic flux $\Phi$, described by the Hamiltonian
\begin{align}
    H(\Phi) &= -t\sum_{j=1}^{L}\sum_{\sigma = \uparrow,\downarrow}\qty(e^{i\Phi/L}c^{\dagger}_{j+1,\sigma}c_{j,\sigma} + \mathrm{h.c.})\nonumber\\
    &+ U\sum_{j=1}^{L}n_{j,\uparrow}n_{j,\downarrow} -\mu\sum_{j=1}^{L}\qty(n_{j,\uparrow} + n_{j,\downarrow}),\label{eq:ham_phi}
\end{align}
where $c_{j,\sigma}$ ($c^\dagger_{j,\sigma}$) annihilates (creates) an electron at site $j$ with spin $\sigma$, $n_{j,\sigma} = c^\dagger_{j,\sigma}c_{j,\sigma}$.
Here, $t>0$ is the nearest-neighbor hopping amplitude, $\mu$ is the chemical potential, and $U\geq 0$ is the on-site interaction strength.
Periodic boundary conditions are imposed.
Hereafter, we set $t=1$.
The Hamiltonian \eqref{eq:ham_phi} conserves the particle number of each spin component,
\begin{equation}
    [H, \hat{N}_\sigma] = 0,
\end{equation}
where $\hat{N}_{\sigma} = \sum_{j}n_{j,\sigma}$ is the total particle number for spin $\sigma$.
Since $[\hat{N}_{\uparrow},\hat{N}_{\downarrow}] = 0$, the Hilbert space decomposes into sectors labeled by the eigenvalues of $\hat{N}_{\uparrow}$ and $\hat{N}_{\downarrow}$. 
Throughout this work, we consider sectors of $\hat{N} = \hat{N}_{\uparrow} + \hat{N}_{\downarrow}$ and $\hat{M} = \hat{N}_{\downarrow}$ with fixed eigenvalues $N$ and $M$, respectively.
In the following, we focus on the zero-magnetization sector ($M=N/2$).

Since the NLDWs are obtained from the flux dependence of the ground-state energy,
we briefly review the Bethe ansatz solution \cite{LiebWu,ShastrySutherland,MartinsFye} in the presence of a magnetic flux.
The energy eigenstates of the Hamiltonian \eqref{eq:ham_phi} are specified by charge quasimomenta $\{k_j\}$ ($j=1,\ldots, N$) and spin rapidities $\{\lambda_l\}$ ($l=1,\ldots,M$) satisfying the Lieb-Wu equations:
\begin{subequations}
    \label{eq:LW}
    \begin{equation}
        k_jL+\Phi + \sum_{l=1}^{M}2\arctan(\frac{\sin{k_j}-\lambda_l}{U/4}) = 2\pi I_j,\label{eq:LW-c}
    \end{equation}
    \begin{eqnarray}
        &&\sum_{j=1}^{N}2\arctan(\frac{\lambda_l-\sin{k_j}}{U/4})\notag\\&&-\sum_{m=1}^{M}2\arctan(\frac{\lambda_l-\lambda_m}{U/2}) = 2\pi J_l,\label{eq:LW-s}
    \end{eqnarray}
\end{subequations}
where the quantum numbers $I_j$ are distinct integers (half-odd integers) for $M$ even (odd) and $J_l$ are integers (half-odd integers) for $N-M$ odd (even).
For the ground state in each sector, the quantum numbers are given by
\begin{equation}
    I_j = \frac{1}{2}\qty(N-2j + s),\quad J_l = \frac{1}{2}\qty(M-2l+1-s')
\end{equation}
with $s = (N-M)~\mathrm{mod}~2$ and $s' = N~\mathrm{mod}~2$.
The corresponding ground-state energy is expressed as
\begin{equation}
    E_0(\Phi;N,M) = -2\sum_{j=1}^{N}\cos{k_j} -\mu N.\label{eq:energy-Bethe}
\end{equation}

\subsection{\label{subsec:nldws}Nonlinear Drude weights}
We next introduce the nonlinear Drude weights (NLDWs).
They are natural higher-order generalizations of the conventional Drude weight and
characterize ballistic transport beyond the linear-response regime \cite{WatanabeOshikawa,WatanabeLiuOshikawa}.
They are defined from the low-frequency limit of the nonlinear optical conductivity.
The $n$th order Drude weight is defined as
\begin{equation}
    D^{(n)} = \lim_{\omega_1,\ldots,\omega_n\to 0}\omega_1\cdots\omega_n\cdot\sigma^{(n)}(\omega_1,\ldots,\omega_n)
\end{equation}
where $\sigma^{(n)}(\omega_1,\ldots,\omega_n)$ is the $n$th-order optical conductivity, introduced in Ref.~\cite{WatanabeOshikawa,WatanabeLiuOshikawa}.
For one-dimensional systems, it can be evaluated from the ground-state energy through the generalized Kohn formula \cite{WatanabeLiuOshikawa,WatanabeOshikawa}:
\begin{equation}
    D^{(n)}= L^{n}\pdv[n+1]{}{\Phi}E_0(\Phi; N, M)\eval_{\Phi = 0}\label{eq:genKohn}.
\end{equation}
The numerical evaluation of Eq.~\eqref{eq:genKohn} is described in the following section.

\section{\label{sec:quarter}Results at quarter-filling}
\subsection{\label{subsec:large-U}Strong-coupling limit}
We first consider the strong-coupling limit,
where the interaction dependence of the NLDWs can be obtained analytically.
In the strong-coupling limit $U\to\infty$,
double occupancy is completely suppressed and charge degrees of freedom are described by free spinless fermions with density $2\nu$ \cite{CaspersIske,OgataShiba,murakami-goehmann-hubbard-t0}.
Consequently, the NLDWs at density $\nu$ are related to those of free spinless fermions at density $2\nu$:
\begin{equation}
    D^{(2n+1)}(U=\infty) = \frac{2}{\pi}(-1)^{n}\sin(\pi\nu)~\label{eq:uinfty}.
\end{equation}
Notably, all NLDWs with odd order have the same magnitude in this limit and differ only by an alternating sign.
The leading correction is given by,
\begin{align}
    &D^{(2n+1)}(U) - D^{(2n+1)}(\infty) \\&= (-4)^{n+2}\frac{e_\mathrm{XXX}}{U}\qty[\frac{\nu\sin(2\pi\nu)}{2\pi} - \frac{\sin^2(\pi\nu)}{\pi^2}] + \mathcal{O}\qty(U^{-2}).\label{eq:largeU}
\end{align}
Here $e_{\mathrm{XXX}}$ is the ground-state energy density of the spin-$\frac{1}{2}$ antiferromagnetic Heisenberg chain,
which is given by $-\ln 2$ in the thermodynamic limit.
The derivation of \eqref{eq:largeU} is presented in Appendix~\ref{app:strong-coupling}.

\subsection{\label{subsec:quarter-num}Numerical results}
\begin{figure*}
    [t]\centering
    \includegraphics[width=0.32\linewidth]{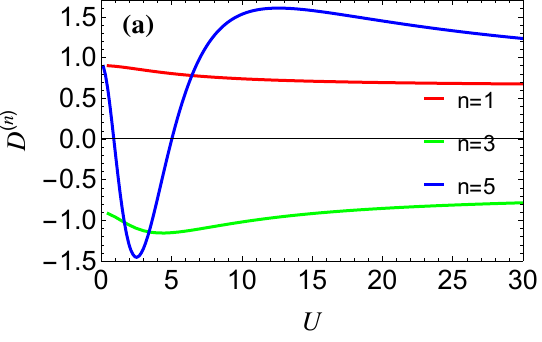}
    \hfill
    \includegraphics[width=0.32\linewidth]{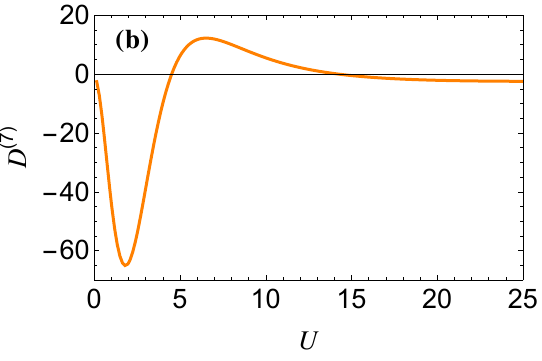}
    \hfill
    \includegraphics[width=0.32\linewidth]{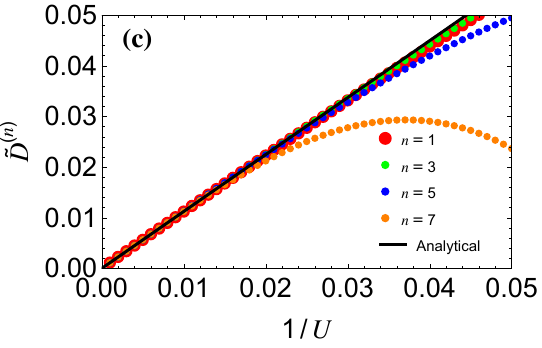}
    \caption{Interaction dependence of the NLDWs at quarter filling for system size $L=204$.
    (a)~First-, third- and fifth-order Drude weights as functions of the interaction strength $U$.
    (b)~Seventh-order Drude weight as a function of $U$.
    (c)~Scaled NLDWs $\tilde{D}^{(n)} = [D^{(n)}(U)-D^{(n)}(\infty)]/(-4)^{(n-1)/2}$ plotted as functions of $1/U$.
    The symbols denote numerical results obtained by solving the Bethe-ansatz equations \eqref{eq:LW}, and the solid line presents the term proportional to $1/U$ in the analytical strong-coupling expansion given by Eq.~\eqref{eq:largeU}.
    }
    \label{fig:NLDW_U}
\end{figure*}

We numerically calculate the first- through seventh-order Drude weights at quarter filling ($N=L/2$) by solving the Bethe ansatz equations \eqref{eq:LW}.
Figure~\ref{fig:NLDW_U} shows the interaction dependence of the NLDWs.
Owing to either inversion or time-reversal symmetry, all even-order Drude weights vanish identically.
Therefore, we focus on the odd-order Drude weights $D^{(1)}$, $D^{(3)}$, and $D^{(5)}$.
As shown in Figs.~\ref{fig:NLDW_U}(a) and \ref{fig:NLDW_U}(b),
all odd-order NLDWs approach finite values in the strong-coupling regime.
In contrast, the interaction dependence becomes highly nonmonotonic in the weak-coupling regime.
In particular, the fifth- and seventh-order Drude weights change sign as the interaction strength increases.
A qualitatively similar behavior has also been reported for NLDWs in the spin-$1/2$ XXZ chain \cite{TanikawaTakasanKatsura},
suggesting that this feature is common to interacting one-dimensional systems.
To examine the strong-coupling behavior in more detail,
we rescale the numerical data according to Eq.~\eqref{eq:largeU}.
As shown in Fig.~\ref{fig:NLDW_U}(c), the 
results for different orders collapse onto a single straight line,
in excellent agreement with the analytical strong-coupling expansion.
This confirms the predicted $1/U$ correction and its 
$n$ dependence given by Eq.~\eqref{eq:largeU}.


\subsection{\label{subsec:cft}Prediction from the low-energy effective theory}
We next compare the numerical results with the prediction of the low-energy effective theory.
To understand the behavior of NLDWs in the thermodynamic limit, we employ the bosonization technique \cite{Giamarchi}. 
Away from half filling,
the low-energy physics of the one-dimensional Hubbard model is described by the spin-charge separated TLL.
Since the magnetic flux couples only to the charge sector,
we consider only the charge Hamiltonian.
Here we follow the discussion in Ref.~\cite{FukusumiBarisic}.
Although their work focused on the XXZ chain, the same arguments straightforwardly apply to the charge part of the Hubbard chain.
Under the renormalization group, it flows to the $c=1$ Gaussian fixed point.
The Hamiltonian of the charge sector is given by
\begin{equation}
    H = H_0 + \delta H,\label{eq:ham_boson}
\end{equation}
where
\begin{equation}
    H_0 = \frac{v}{2\pi}\int_{0}^{La}dx\,\qty[\frac{1}{K}:\qty(\frac{d \phi}{d x})^2: + K:\qty(\frac{d \theta}{d x})^2:]
\end{equation}
is the free-boson Hamiltonian, and $\delta H$ consists of irrelevant perturbations \cite{EsslerPereiraSchneider}.
Here, $v$ is the charge velocity, $a$ is the lattice constant, and $K$ is the TLL parameter.
We set $x = na$ ($n=1,2,\ldots, L$) and 
take the continuum limit $L\to \infty$, $a\to 0$ with $La$ kept fixed.
The bosonic fields $\phi(x)$ and $\theta(x)$ 
satisfy the commutation relation
\begin{equation}
    [\phi(x),\partial_y\theta(y)] = i\pi\delta(x-y).
\end{equation}
The flux dependence is incorporated 
by shifting the dual field as \cite{Loss,Schmeltzer,SchmeltzerBerkovits}
\begin{equation}
    \theta^{\Phi}(x) = \theta(x) - \frac{\sqrt{2}}{La}\Phi x,
\end{equation}
which corresponds to the twisted boundary condition of the underlying fermions. 
This shift can be absorbed into the zero mode of the dual field. Consequently,
the ground-state energy of $H_0$ is given by
\begin{equation}
    E^{(0)}_0 = \epsilon_0 L - \frac{\pi vc}{6La} + \frac{2\pi v}{La}\frac{K}{2}\qty(\frac{\Phi}{\pi})^2,\label{eq:egs-finite}
\end{equation}
where $c = 1$ is the central charge and $\epsilon_0$ is the ground-state energy density of the lattice Hamiltonian in the thermodynamic limit.
The derivation of this finite-size expression is given in Appendix~\ref{app:Umklapp}.

Since the flux dependence of $E^{(0)}_0$ is purely quadratic, the higher-order Drude weights originate entirely from the irrelevant perturbations $\delta H$. 
The leading irrelevant operators originating from band curvature \cite{Haldane1,Lukyanov, Cardy1, LukyanovTerras,BortzKarbachSchneiderEggert,EsslerPereiraSchneider} gives a finite contribution to the third-order Drude weight.
Higher-order band-curvature operators similarly give rise to finite contributions to higher-order NLDWs, with corrections of order $\mathcal{O}(L^{-2})$.

In addition to the band-curvature terms, there exists another important irrelevant perturbation.
At quarter filling, the Umklapp term
\begin{equation}
    \delta H_\mathrm{Umklapp} = a^{8K-1}\lambda_U\int_{0}^{La}\frac{dx}{2\pi}2:\cos(4\sqrt{2}\phi):,
\end{equation}
which transfers four electrons between the two Fermi points,
is commensurate \cite{YoshiokaTsuchiizuSuzumura, Nakamura, OtsukaNakamura, Giamarchi2}.
The scaling dimension of the Umklapp operator is $8K$.
Since the first-order correction to the ground-state energy due to this term vanishes, the leading contribution arises at second order in $\lambda_U$, which leads to
the finite-size correction to 
$E^{(0)}_0$,
\begin{equation}
    \delta E_0 = -\frac{\lambda_U^2a}{4Kv}\qty(\frac{2\pi}{L})^{16K-3}\qty(\frac{1}{2+\frac{\Phi}{\pi}} + \frac{1}{2-\frac{\Phi}{\pi}}).
\end{equation}
Thus, for odd $n$, the contribution of the Umklapp term to the $n$th-order Drude weight, $D^{(n)}_{\mathrm{Umklapp}}$, 
scales as
\begin{equation}
    D^{(n)}_\mathrm{Umklapp}\sim L^{n+3-16K}\label{eq:scale-D}.
\end{equation}
Therefore, the Umklapp contribution diverges in the thermodynamic limit when
\begin{equation}
    n > 16K-3 \label{eq:criterion}.
\end{equation}
The details of the perturbative calculation are presented in Appendix~\ref{app:Umklapp}.


The criterion \eqref{eq:criterion} depends on the TLL parameter $K$, whose value varies continuously with the interaction strength.
We therefore determine $K$ from the Bethe ansatz through the exact relation between the Drude weight and compressibility \cite{Haldane2,SanoOno,LaflorencieCapponiSorensen} and identify the parameter regions 
where the Umklapp contribution is expected to diverge.
\begin{figure}[h]
    \centering
    \includegraphics[width=0.48\linewidth]{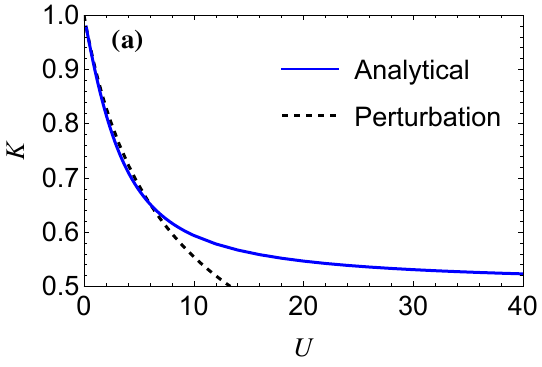}
    \hfill
    \includegraphics[width=0.48\linewidth]{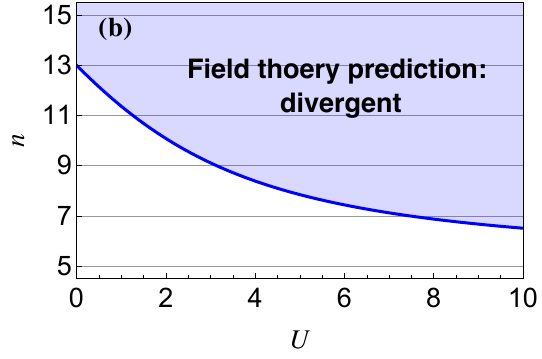}
    \caption{TLL parameter at quarter filling and the corresponding criterion for the divergence of NLDWs.
    (a) Interaction dependence of $K$ obtained from the Bethe ansatz solution.
    The solid curve represents the numerical result, while the dashed curve shows the weak-coupling perturbative result.
    (b) Threshold value of the order, $16K-3$, as a function of the interaction strength.
    The shaded region indicates the parameter regime where the Umklapp contribution to the NLDWs is expected to diverge according to Eq.~\eqref{eq:criterion}.}
    \label{fig:K}
\end{figure}
Figure~\ref{fig:K}(a) shows the interaction dependence of $K$ at quarter filling, together with its weak-coupling asymptotic expansion. The TLL parameter satisfies $1/2 \le K \le 1$ 
\cite{FrahmKorepin}. Since no divergence of NLDWs occurs exactly at $K=1/2$ ($U=\infty$) and $K=1$ ($U=0$), both of which reduce to free fermions, we restrict ourselves to the case where $1/2 < K < 1$.

Figure~\ref{fig:K}(b) shows the corresponding interaction dependence of $16K-3$, which determines the criterion for the divergence of the Umklapp contribution according to Eq.~\eqref{eq:criterion}.
In the strong-coupling regime limit ($K\simeq 1/2$), the divergence is predicted to first appear at $n=7$,
whereas in the weak-coupling regime $(K\simeq 1)$ it first appears at $n=13$.
Therefore, the low-energy effective field theory predicts divergent NLDWs over an increasingly broad range of interaction strength as the order is increased.

\subsection{\label{subsec:diverge}Absence of divergence in NLDWs at quarter-filling}
\begin{figure*}[t]
    \centering
    \includegraphics[width=\linewidth]{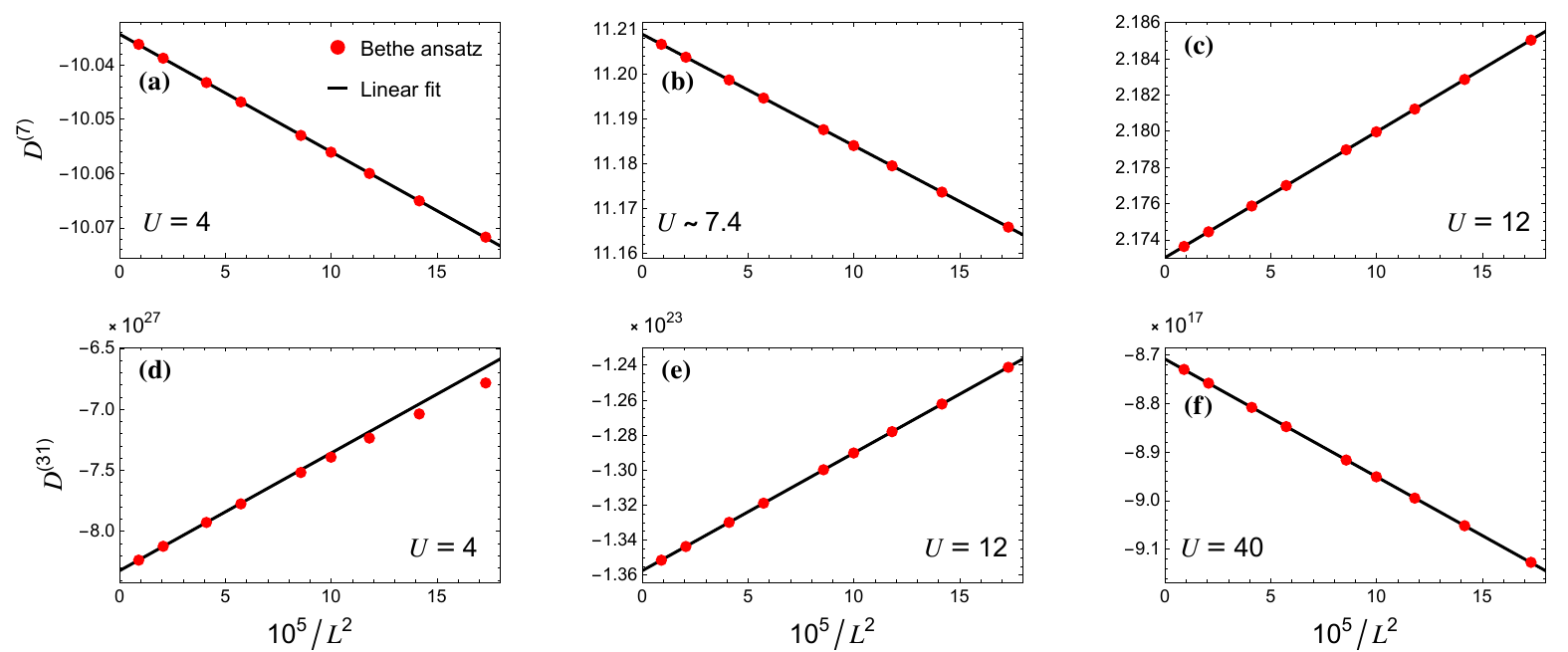}
    \caption{Finite-size scaling of the $n$th-order Drude weights $D^{(n)}$ at quarter filling.
    The upper and lower panels show $D^{(7)}$ and $D^{(31)}$, respectively.
    For $n=7$, the interaction strengths are chosen below, close to, and above the threshold $16K-3$ predicted by the low-energy effective theory, as shown in Fig.~\ref{fig:K}.
    For $n=31$, field theory predicts that $D^{(n)}$ diverges for all interaction strengths.
    In all cases, the data are well described by linear functions of $1/L^2$, indicating that the nonlinear Drude weights remain finite in the thermodynamic limit.}
    \label{fig:NLDWs_nlarge}
\end{figure*}

The field-theoretical prediction can now be compared with the numerical Bethe-ansatz results.
Figure~\ref{fig:NLDWs_nlarge} shows the finite-size scaling of $D^{(7)}$ and $D^{(31)}$ for representative interaction strength spanning both sides of the threshold predicted by Eq.~\eqref{eq:criterion}.
All calculations were performed in {\it Mathematica} with the numerical precision set sufficiently high, so that numerical errors are negligible.
Contrary to the prediction of the low-energy effective theory, the numerical results exhibit $\mathcal{O}(L^{-2})$ finite size corrections.
Therefore, exact results 
from the Bethe ansatz are not consistent with the predictions 
of the field-theoretical analysis.
This discrepancy indicates that the reason for the absence of the predicted divergence requires further investigation.
Possible explanations,
including the 
potential absence of the corresponding Umklapp interaction and limitations of the perturbative field-theoretical treatment,
will be discussed in Sec.~\ref{sec:discussion}.

\section{\label{sec:scaling}Scaling of NLDWs at metal-insulator transition}
We now turn to the half-filled case, where the ground state is a Mott insulator.
Unlike quarter filling, all NLDWs vanish in the thermodynamic limit.
Nevertheless, finite-size systems exhibit pronounced size dependence,
especially in the weak-coupling regime.
In this section, we first clarify the origin of this behavior and then discuss the scaling properties of the NLDWs near the metal-insulator transition.

\subsection{\label{sec:half}Finite-size scaling of NLDWs at half-filling}
\begin{figure}[h]
\centering\includegraphics[width=0.9\linewidth]{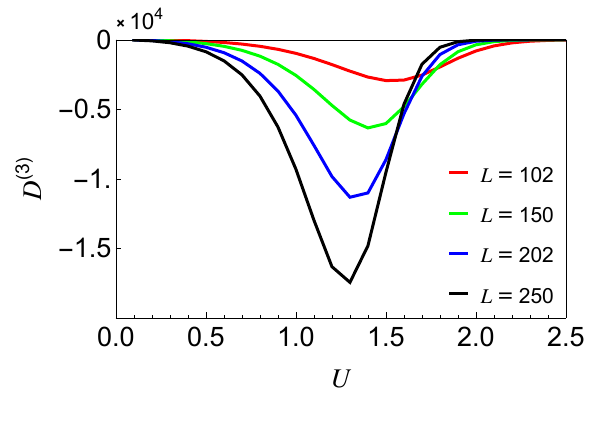}
    \caption{Third-order Drude weight at half filling as a function of the interaction strength for several system sizes.
    Although the Drude weight vanishes in the thermodynamic limit,
    pronounced finite-size effects appear in the weak-coupling regime.}
    \label{fig:half_D3}
\end{figure}

\begin{figure}[h]
    \centering
    \includegraphics[width=0.9\linewidth]{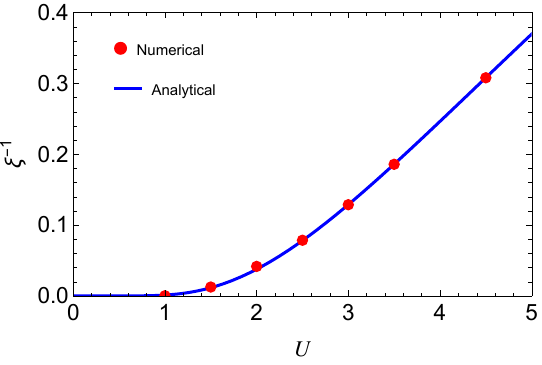}
    \caption{Inverse correlation length at half filling.
    The solid blue curve represents the 
    analytical result given by Eq.~\eqref{eq:xi}, while the symbols are extracted from the finite-size scaling of the third-order Drude weight using Eq.~\eqref{eq:scaling-n}.}
    \label{fig:corleng}
\end{figure}

Figure~\ref{fig:half_D3} shows the third-order Drude weight at half filling for several system sizes, obtained numerically from the Bethe ansatz equations.
At half filling, the Drude weight changes sign between the $L=4k$ and $L=4k+2$ sequences \cite{FyeMartinsScalapinoetal,StaffordMillis,StaffordMillisShastry}.
Since this sign alternation does not affect the finite-size scaling discussed below, we restrict ourselves to the $L=4k+2$ sequence throughout this work.
Although all NLDWs vanish in the thermodynamic limit owing to the Mott gap,
the numerical data exhibit a pronounced increase with system size in the weak-coupling regime.
This apparent contradiction is resolved by the asymptotic finite-size scaling.
The asymptotic behavior can be obtained by extending the analysis of the linear Drude weight \cite{StaffordMillis} to arbitrary odd orders:
\begin{equation}
  D^{(2n+1)} \sim (-1)^{n}L^{2n+1/2}\exp[-L/\xi(U)].
  \label{eq:scaling-n}
\end{equation}
Here, the correlation length $\xi(U)$ is obtained 
as
\begin{equation}
    \frac{1}{\xi(U)} = \frac{1}{4}\int_{1}^{\infty}dy~\frac{\ln(y+\sqrt{y^2-1})}{\cosh(2\pi y/U)}\label{eq:xi}
\end{equation}
and the prefactor depends only on $U$.
The derivation of Eq.~\eqref{eq:xi} and an analytical expression for the prefactor are presented in Appendix~\ref{app:half-filling}.
The inverse correlation length $1/\xi$ is shown in Fig~\ref{fig:corleng}, demonstrating 
excellent agreement 
between the analytical expression and the numerical results.
For weak interactions, the correlation length becomes much larger than the accessible system sizes.
Consequently, the exponential suppression is not yet effective, leading to the pronounced finite-size dependence observed in the numerical results.

\subsection{\label{sec:trans}Scaling behavior of NLDWs near the critical point}
\begin{figure*}
    \centering
    \includegraphics[width=\linewidth]{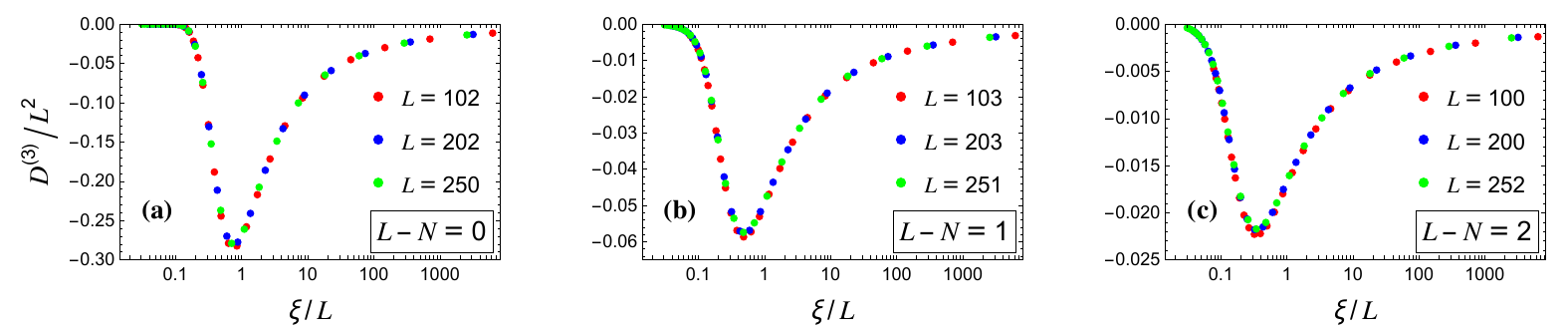}
    \caption{Verification of the hyperscaling ansatz \eqref{eq:hyperscaling_n} for the third-order Drude weight.
    The scaled quantity $D^{(3)}/L^2$ is plotted as a function of $\xi/L$ for (a) $L-N=0$, (b) $L-N=1$, (c) $L-N=2$.
    Different symbols correspond to different system sizes.
    For each fixed value of $L-N$, the numerical data collapse onto a single curve, supporting the proposed hyperscaling form.}
    \label{fig:hyperscalingD3}
\end{figure*}

In Ref.~\cite{StaffordMillis}, it 
has been verified that the behavior of the linear Drude weight in the vicinity of the metal-insulator critical point ($U=0$, $\nu=N/L= 1$) obeys the hyperscaling form \cite{KimWeichman}
\begin{equation}
    D^{(1)}(\nu,L,U) = Y^{(1)}(\xi\delta,\xi/L),
\end{equation}
where $Y^{(1)}(x,y)$ is a scaling function, the correlation length $\xi$ is given in Eq.~\eqref{eq:xi}, and $\delta = 1-\nu$ is the doping density.
This scaling form naturally suggests a generalization to higher-order NLDWs.
We extend this hyperscaling form to 
the nonlinear case and propose an ansatz for the $n$th Drude weight near the critical point:
\begin{equation}
    D^{(n)}(\nu,L,U)/L^{n-1} \sim Y^{(n)}(\xi\delta,\xi/L).\label{eq:hyperscaling_n}
\end{equation}
To verify this scaling form numerically, we fix the number of doped holes $L-N$.
Since $\delta = (L-N)/L$, the first scaling variable becomes
\begin{equation}
    \xi\delta = (L-N)\frac{\xi}{L}.
\end{equation}
Therefore, for each fixed value of $L-N$, the scaling function depends only on the single variable $\xi/L$.
Figure~\ref{fig:hyperscalingD3} shows the scaling collapse of the third-order Drude weight.
For each fixed value of $L-N$, data obtained for different interaction strengths and system sizes collapse onto a single curve when plotted against $\xi/L$.
This excellent collapse confirms the proposed hyperscaling ansatz near Mott transition.

\section{\label{sec:discussion}Discussion and Outlook}

In this work, we have investigated the nonlinear Drude weights (NLDWs) in the one-dimensional Hubbard model and clarified their behavior both at quarter filling and in the vicinity of the Mott transition.
In particular, we found no evidence for the divergences predicted by the low-energy effective theory at quarter filling,
while extending the scaling theory of Stafford and Millis \cite{StaffordMillis} to higher-order responses and establishing the corresponding exponential and hyperscaling behaviors near the Mott transition.
These results establish the NLDWs as a useful quantity for investigating nonlinear transport in the one-dimensional Hubbard model.

The absence of the expected divergences at quarter filling raises an interesting question regarding the role of Umklapp interactions in nonlinear transport.
Although higher-order Umklapp processes are allowed by the lattice symmetry at quarter filling,
the corresponding divergences are not observed in our numerical calculations even at sufficiently high orders.
One might think that the absence of the divergence is specific to the zero-flux point,
where a special cancellation may occur.
Since the NLDWs are obtained from derivatives of the ground-state energy density with respect to the magnetic flux evaluated at $\Phi=0$,
such a possibility cannot be excluded a priori. 
However, we show in Appendix~\ref{app:phi} that the absence of the divergence persists at finite $\Phi$, indicating that it is not a consequence of such a cancellation.
Another possible explanation is that the coupling constant of the corresponding Umklapp operator may vanish identically because of additional constraints beyond those imposed by lattice symmetry,
possibly related to the special symmetries of the one-dimensional Hubbard model \cite{Shastry,Fukai}.
Such a mechanism, however, is unlikely to be a consequence of integrability alone,
since analogous divergences induced by Umklapp interactions are found in the integrable spin-$\frac12$ XXZ 
chain \cite{WatanabeOshikawa,TanikawaKatsura,TanikawaTakasanKatsura}.
Alternatively, the coupling constant may be nonzero but sufficiently small that the associated divergent behavior is not accessible within the system sizes and orders considered here.
Finally, the perturbative calculation based on the low-energy effective field theory requires further examination when applied to higher-order nonlinear responses.
Clarifying the origin of this discrepancy remains an important open problem.

It is also worth noting that we find a similar situation in the spin-$\frac{1}{2}$ XXZ chain,
where the expected divergences are absent despite the presence of symmetry-allowed Umklapp interactions at quarter filling (see Appendix~\ref{app:xxz}).
This observation suggests that the discrepancy between the predictions of the low-energy effective theory and the Bethe-ansatz results may not be specific to the Hubbard model but could instead reflect a more general limitation of the current effective field theory description of 
nonlinear transport.
Further investigations from both field-theoretical and microscopic viewpoints will therefore be desirable.

The present work also opens several directions for future research. 
An immediate extension is to the extended Hubbard model, in which higher-order Umklapp interactions are known to drive the Mott transition at commensurate fillings \cite{MilaZotos, PencMila, ShirakawaJeckelmann, Nakamura, EjimaGebhardNishimoto}.
Investigating NLDWs in this model would therefore provide an ideal testing ground for the role of higher-order Umklapp interactions in nonlinear transport.
Another promising direction is to investigate the polarization amplitude, which has 
been proposed as a probe of Umklapp interactions \cite{KobayashiNakagawaFukusumiOshikawa,FuruyaNakamura}.
Since the present work suggests the possible absence of the corresponding Umklapp interactions at quarter filling,
it would be interesting to examine whether the polarization amplitude can provide independent evidence for this scenario.
Extending the present analysis to finite temperatures is another important problem \cite{FujimotoKawakami, LuoBasakPuGuan,UrichukKlumperSirker,FabaBiswasGopalakrishnanVasseurParameswaran}, where the interplay between thermal fluctuations and nonlinear transport remains largely unexplored.

We hope that the present work will stimulate further studies of nonlinear transport in interacting electron systems and contribute to a deeper understanding of the role of Umklapp interactions in nonlinear responses.

\begin{acknowledgments}
We would like to thank Yoshiki Fukusumi and Kazuaki Takasan for helpful discussions. 
T.I. was supported by Forefront Physics and Mathematics Program to Drive Transformation (FoPM), a World-leading
Innovative Graduate Study (WINGS) Program, the University of Tokyo. 
H.K. was supported by JSPS KAKENHI Grants No. JP23K25783 and No. 23K25790. 
\end{acknowledgments}

\section*{data availability}
The data that support the findings of this article are openly available \cite{zenodo}.

\appendix

\section{Strong-coupling expansion\label{app:strong-coupling}}
In this Appendix, we derive the strong-coupling expansion of the NLDWs following the arguments of Ref.~\cite{Essler}.
We first derive the limit of $U\to \infty$, where the one-dimensional Hubbard model reduces to free spinless fermions \cite{CaspersIske,OgataShiba},
and then obtain the leading correction in $1/U$ from the Lieb-Wu equations \eqref{eq:LW}.

\subsection{Infinite-$U$ limit}
For simplicity, we consider the sector with even $N$, odd $M$, and zero magnetization ($M=N/2$).
The extension to finite magnetization is straightforward.
The other parity sectors can be treated similarly and lead to the same 
results in the thermodynamic limit.

In the large $U$ limit, 
the spin rapidities are of order $U$, whereas the charge quasimomenta remain finite.
Introducing $\Lambda_l = \lambda_l/U$ and taking the limit $U\to \infty$ in Eqs.~\eqref{eq:LW-c} and \eqref{eq:LW-s}, we obtain
\begin{align}
    &e^{ik_jL + i\Phi} = (-1)^M\prod_{l=1}^{M}\frac{i/4-\Lambda_l}{i/4+\Lambda_l},\label{eq:LW-largeU-c}\\
    &\qty(\frac{i/4 + \Lambda_l}{i/4-\Lambda_l})^N = (-1)^{N-M-1}\prod_{m\neq l}\frac{i/2 + \Lambda_l-\Lambda_m}{i/2 - \Lambda_l +\Lambda_m}. \label{eq:LW-largeU-s}
\end{align}
Taking the product over all $l$ in Eq.~\eqref{eq:LW-largeU-s} gives
\begin{equation}
    \qty(\prod_{l=1}^{M}\frac{i/4 + \Lambda_l}{i/4-\Lambda_l})^N = 1,
\end{equation}
which implies that the right-hand side of Eq.~\eqref{eq:LW-largeU-c} is equal to $e^{i2\pi m/N}$ with $m=0,\ldots,N-1$.
The quasimomenta are therefore
\begin{equation}
    k_j = \frac{2\pi I_j}{L} + \frac{2\pi m}{LN} - \frac{\Phi}{L},\label{eq:kj-uinfty}
\end{equation}
where $I_j$ are distinct half-integers.
For sufficiently small $\Phi$, the ground-state quantum numbers are given by
\begin{equation}
    I_j = \frac{-N-1+2j}{2},\quad j=1,\ldots,N,
\end{equation}
with $m=0$, which gives
\begin{equation}
    E_0(\Phi;N,M) = -2\frac{\sin(N\pi/L)}{\sin(\pi/L)}\cos(\frac{\Phi}{L}).
\end{equation}
This is precisely the ground-state energy of free spinless fermions with particle density $2\nu$.
Therefore, the $(2n+1)$th-order Drude weight of the Hubbard model, $D^{(2n+1)}_\mathrm{Hubbard}$, is related to that of free spinless fermions, $D^{(2n+1)}_\mathrm{SF}$ by
\begin{equation}
    D_\mathrm{Hubbard}^{(2n+1)}(\nu,U=\infty) = D_\mathrm{SF}^{(2n+1)}(2\nu).
\end{equation}
Here, the filling is doubled because the two spin species of the Hubbard model are mapped onto a single species of spinless fermions in the charge sector.
This immediately gives Eq.~\eqref{eq:uinfty}.

\subsection{Leading $1/U$ correction}
We next derive the leading correction in $1/U$.
The Bethe roots are expanded as
\begin{equation}
\begin{aligned}
    k_j &= k_j^{(0)} + \frac{1}{U}\Delta k_j + \mathcal{O}\qty(U^{-2})\\
    \lambda_l &= U\Lambda_l + \Delta\Lambda_l + \mathcal{O}\qty(U^{-2}),
    \end{aligned}
    \label{eq:expand-root}
\end{equation}
where $k_j^{(0)}$ is identified with $k_j$ in Eq.~\eqref{eq:kj-uinfty}, and $\Lambda_l$ satisfies Eq.~\eqref{eq:LW-largeU-s}.
Substituting Eq.~\eqref{eq:expand-root} into the Lieb-Wu equations \eqref{eq:LW} and collecting terms of order $U^{-1}$, we obtain
\begin{align}
  &\frac{1}{4}\Delta k_jL = 2\sum_{l=1}^{M}\frac{\Delta\Lambda_l - \sin{k_j^{(0)}}}{16\Lambda_l^2+1}\label{eq:1/u-1-c},\\
  &\sum_{j=1}^{N}\frac{\Delta\Lambda_l - \sin{k_j^{(0)}}}{16\Lambda_l^2 + 1} = 2\sum_{m=1}^{M}\frac{\Delta\Lambda_l-\Delta\Lambda_m}{16(\Lambda_l+\Lambda_m)^2 + 4}\label{eq:1/u-1-s}.
\end{align}
Summing Eq.~\eqref{eq:1/u-1-s} over $l$ gives
\begin{equation}
     \sum_{l=1}^{M}\frac{\Delta\Lambda_l}{16\Lambda_l^2+1} = \frac{1}{N}\sum_{j=1}^{N}\sum_{l=1}^{M}\frac{\sin{k_j^{(0)}}}{16\Lambda_l^2+1}.
\end{equation}
Substituting this relation into Eq.~\eqref{eq:1/u-1-c}, we obtain
\begin{equation}
\begin{aligned}
  \frac{1}{4}\Delta k_jL &= {2}\sum_{l=1}^{M}\frac{1}{16\Lambda_l^2+1}\qty(\frac{1}{N}\sum_{n=1}^{N}\sin{k_n^{(0)}}-\sin{k_j^{(0)}})\\
  &= e_\mathrm{XXX}\sum_{n=1}^{N}\qty(\sin{k_j^{(0)}} - \sin{k_n^{(0)}}),
  \end{aligned}
\end{equation}
where
\begin{equation}
    e_\mathrm{XXX} = -\frac{2}{N}\sum_{l}^{M}\frac{1}{16\Lambda_l^2+1} \to -\ln2\quad (N\to\infty)
\end{equation}
is the ground-state energy density of the spin-$1/2$ antiferromagnetic Heisenberg chain in the thermodynamic limit~\cite{CNYangCPYang1,CNYangCPYang2}.

Substituting the corrected quasimomenta into 
Eq. \eqref{eq:energy-Bethe}, the ground-state energy up to order $1/U$ becomes
\begin{equation}
\begin{aligned}
    &E_0(\Phi;N,M) = -2\frac{\sin(N\pi/L)}{\sin(\pi/L)}\cos(\frac{\Phi}{L}) \\
    &+ \frac{8e_\mathrm{XXX}}{U}\qty[\frac{N}{L}\sum_{j=1}^{N}\qty(\sin{k_j^{(0)}})^2 - \frac{1}{L}\qty(\sum_{j=1}^{N}\sin{k_j^{(0)}})^2].
    \end{aligned}
\end{equation}
Differentiating the ground-state energy with respect to the magnetic flux and taking the thermodynamic limit yields
\begin{equation}
    \begin{aligned}
        &D^{(2n+1)} = (-1)^n\frac{2}{\pi}\sin(\pi\nu) \\&+ (-4)^{n+2}\frac{e_\mathrm{XXX}}{U}\qty[\frac{\nu\sin(2\pi\nu)}{2\pi} - \frac{\sin^2(\pi\nu)}{\pi^2}] + \mathcal{O}(U^{-2}),
    \end{aligned}
\end{equation}
which is Eq.~\eqref{eq:largeU} in the main text.

\section{Umklapp contribution to the ground-state energy\label{app:Umklapp}}
In this Appendix, we derive the contribution of the Umklapp interaction to the ground-state energy within perturbed CFT.
The essential point is that the external magnetic flux shifts the zero mode of the Gaussian CFT \cite{Loss,Schmeltzer,SchmeltzerBerkovits}.
As a consequence, the energy differences between the ground state and the intermediate states connected by the Umklapp operator become flux dependent,
which leads to the singular flux dependence responsible for the divergence of the NLDWs.

\subsection{Effective Hamiltonian and flux dependence}
We treat the Umklapp interaction perturbatively around the Gaussian fixed point:
\begin{align}
        &H = H_0 + \delta H_\mathrm{Umklapp},\\
        &H_0 = \frac{v}{2\pi}\int_{0}^{La}dx\qty[\frac{1}{K}:\qty(\dv{\phi}{x})^2: + K:\qty(\dv{\theta}{x})^2:],\\
        &\delta H_\mathrm{Umklapp} = a^{\Delta-1}\lambda_U \int_0^{La}\frac{dx}{2\pi}2:\cos(4\sqrt{2}\phi):.\label{eq:Umklapp}
\end{align}
Here, $\Delta = 8K$ is the scaling dimension of the perturbing operator and the coupling constant $\lambda_U$ is normalized to have the dimension of energy.
By rescaling the bosonic fields as
\begin{equation}
    \tilde{\phi}(x) = \frac{1}{\sqrt{K}}\phi(x),\quad \tilde{\theta}(x) = \sqrt{K}\theta(x),
\end{equation}
the Hamiltonian becomes
\begin{align}
    &H_0 = \frac{v}{2\pi}\int_{0}^{La}dx\qty[:\qty(\dv{\tilde{\phi}}{x})^2: + :\qty(\dv{\tilde{\theta}}{x})^2:],\\
    &\delta H_\mathrm{Umklapp} = a^{\Delta -1}\lambda_U\int_{0}^{La}\frac{dx}{2\pi}2:\cos(4\sqrt{2K}\tilde{\phi}):.
\end{align}
The mode expansions of these fields are performed as
\begin{align}
    &\tilde{\phi}(x) = -\frac{i\pi}{La}\sum_{n\neq 0}\frac{e^{-\alpha|p_n|/2 +ip_nx}}{p_n}\qty(a_{n}+\bar{a}_{-n}) \\
    &\qquad + \frac{\pi x}{La}Q + \phi_0,\\
    &\tilde{\theta}(x) = \frac{i\pi}{La}\sum_{n\neq 0}\frac{e^{-\alpha|p_n|/2+ip_n x}}{p_n}\qty(a_n-\bar{a}_{-n})\\
    &\qquad - \frac{\pi x}{La}Q' - \theta_0,
\end{align}
where $p_n = 2\pi n/La$ and mode operators satisfy
\begin{equation}
    [a_n, a_m] = [\bar{a}_n,\bar{a}_m] = n\delta_{n+m,0},\quad [a_n,\bar{a}_m] = 0
\end{equation}
and
\begin{equation}
    [\phi_0,Q'] = [\theta_0,Q] = -i.\label{eq:com-zeromodes}
\end{equation}
Here, $\alpha$ is an ultraviolet cutoff of the order of the lattice constant.
Then, $H_0$ is diagonalized as
\begin{equation}
    H_0 = \frac{2\pi v}{La}\qty[\sum_{n>0}\qty(a_{-n}a_n + \bar{a}_{-n}\bar{a}_{n}) + \frac{1}{4}\qty(Q^2 + Q'^2)].
\end{equation}
The zero modes $Q$ and $Q'$ are related to the changes in the particle numbers at the two Fermi points.
We denote the corresponding integer quantum numbers for spin $\sigma$ at the right (left) Fermi point by $N_{R,\sigma}$ ($N_{L,\sigma}$).
The zero modes are quantized as \cite{CalabreseEsslerLauchli}
\begin{align}
    Q &= \frac{1}{\sqrt{2K}}(N_{R,\uparrow}+N_{R,\downarrow} + N_{L,\uparrow} + N_{L,\downarrow}),\\
    Q' &= \sqrt{\frac{K}{2}}\qty(N_{R,\uparrow}+N_{R,\downarrow} - N_{L,\uparrow} - N_{L,\downarrow}).
\end{align}
In general, the zero-mode quantum numbers also contain independent contributions from the spin degrees of freedom.
In the following, we restrict ourselves to the spin-symmetric sector, $N_{R,\uparrow}=N_{R,\downarrow}$ and $N_{L,\uparrow} = N_{L,\downarrow}$.
This is sufficient for the present purpose because the spin contributions to the ground-state energy are independent of the magnetic flux $\Phi$, and hence do not contribute to the NLDWs.
The magnetic flux is introduced by imposing a twisted boundary condition on the underlying fermions. Using the standard bosonization formula for the fermion operators \cite{Giamarchi,CalabreseEsslerLauchli,Schmeltzer}, this twist is implemented by shifting the dual bosonic field as
\begin{equation}
    \tilde{\theta}^\Phi(x) = \tilde{\theta}(x) - \frac{\sqrt{2K}}{La}\Phi x.
\end{equation}
The resulting flux dependence can be absorbed into the zero mode $Q'$ \cite{Loss,Schmeltzer,SchmeltzerBerkovits}.
This flux-induced shift may be viewed as a spectral flow in the $U(1)$ sector \cite{SchwimmerSeiberg}.
The zero modes are then written as 
\begin{equation}
    Q^\Phi = \frac{\mathcal{N}}{\sqrt{2K}},\quad Q'^\Phi = \sqrt{{2K}}\qty(\mathcal{J}+\frac{\Phi}{\pi}),
\end{equation}
where
\begin{align}
    \mathcal{N} &= N_{R,\uparrow}+N_{L,\uparrow} + N_{R,\downarrow} + N_{L,\downarrow}\\
    \mathcal{J} &= \frac{N_{R,\uparrow}+N_{R,\downarrow} - N_{L,\uparrow} - N_{L,\downarrow}}{2}.
\end{align}
Here, the integer $\mathcal{N}$ characterizes changes in the total particle number,
while $\mathcal{J}$ characterizes the transfer of particles between the two Fermi points.

The primary states of the Gaussian CFT are defined as the eigenstates of $Q$ and $Q'$ that are annihilated by $a_n$ and $\bar{a}_n$ for $n > 0$.
They are written explicitly as
\begin{equation}
    \ket{\mathcal{N},\mathcal{J}} = \exp[-i\frac{\mathcal{N}}{\sqrt{2K}}\theta_0 - i\sqrt{2K}\mathcal{J}\phi_0]\ket{0},
\end{equation}
where the vacuum $\ket{0}$ is defined by $a_n\ket{0} = \bar{a}_n\ket{0} = 0$ for $n>0$ and $Q\ket{0} = Q'\ket{0} = 0$.
The corresponding eigenenergies are
\begin{equation}
    E^{(0)}_{\mathcal{N},\mathcal{J}}(\Phi) = \frac{2\pi v}{La}\qty[\frac{\mathcal{N}^2}{8K} + \frac{K}{2}\qty(\mathcal{J}+\frac{\Phi}{\pi})^2] - \frac{\pi vc}{6La}\label{eq:nonperturb},
\end{equation}
where the last term is the universal finite-size correction to the ground-state energy arising from conformal invariance \cite{BloteCardyNightingale}.
In the zero-magnetization sector,
the ground state is given by $\ket{0,0}$ for $-\pi < \Phi < \pi$.
The ground-state energy therefore takes the finite-size form shown in Eq.~\eqref{eq:egs-finite}.
The excited states consist of the primary states other than $\ket{0,0}$,
together with the descendant states obtained by acting with $a_{-n}$ and $\bar{a}_{-n}$ ($n>0$) on any primary state.


\subsection{Perturbed CFT calculation of the Umklapp contribution}
It follows from the commutation relation Eq.~\eqref{eq:com-zeromodes} that
the zero-mode part of the Umklapp operator in Eq.~\eqref{eq:Umklapp} shifts $Q'$ by $\pm 4\sqrt{2K}$, or equivalently, $\mathcal{J}\to \mathcal{J}\pm 4$.
Consequently, the first-order correction to the ground-state energy from the Umklapp interaction vanishes:
\begin{equation}
    E_0^{(1)} = \mel{0,0}{\delta H_\mathrm{Umklapp}}{0,0} = 0.
\end{equation}
The leading correction therefore appears at second order,
\begin{equation}
    E_0^{(2)} = -\sum_{\alpha\neq 0}\frac{\qty|\mel{\alpha}{\delta H_\mathrm{Umklapp}}{0}|^2}{E_\alpha^{(0)} - E_0^{(0)}} \label{eq:second-order},
\end{equation}
which can be evaluated within perturbed conformal field theory following Refs.~\cite{AlcarazBarberBatchelor, FukusumiBarisic}.
Here, we denote the complete set of eigenstates of $H_0$ by $\{\ket{\alpha}\}$, with $E_\alpha^{(0)}$ being the corresponding unperturbed energies, and $\ket{0}$ representing the ground state $\ket{0,0}$.
The relevant intermediate primary states are $\ket{0,\pm4}$.
Although descendant states also contribute to the second-order correction in principle,
their contributions are subleading in the large-$L$ expansion.
We therefore retain only $\ket{0,\pm4}$.
Their excitation energies relative to the ground state are
\begin{equation}
    E^{(0)}_{0,\pm4}(\Phi) - E^{(0)}_{0,0}(\Phi) = \frac{2\pi v}{La}4K\qty(2\pm\frac{\Phi}{\pi}). \label{eq:excited-energy}
\end{equation}
The corresponding matrix elements are obtained 
using the standard results from CFT \cite{Cardy1}:
\begin{align}
&\mel{\mathcal{N},\mathcal{J}}{\delta H_\mathrm{Umklapp}}{0,0} \\
    &= \lambda_U\delta_{\mathcal{N},0}\qty(\delta_{\mathcal{J},4} + \delta_{\mathcal{J},-4})\qty(\frac{2\pi}{L})^{\Delta-1}.\label{eq:Cardy}
\end{align}
Using Eqs.~\eqref{eq:second-order}, \eqref{eq:excited-energy}, and \eqref{eq:Cardy},
we obtain
\begin{equation}
    E_0^{(2)} = -\frac{\lambda_U^2a}{4Kv}\qty(\frac{2\pi}{L})^{16K-3}\qty(\frac{1}{2+\frac{\Phi}{\pi}} + \frac{1}{2-\frac{\Phi}{\pi}}).
\end{equation}
Differentiating this expression with respect to the flux gives the finite size scaling of the $n$th-order Drude weight in Eq.~\eqref{eq:scale-D}. 

\section{Finite-size scaling at half filling\label{app:half-filling}}
In this Appendix, we derive the asymptotic behavior of the NLDWs at half filling in the large-$L$ regime.
Our derivation follows the analysis of the linear Drude weight in Ref.~\cite{StaffordMillis}, which we extend to NLDWs by differentiating the finite-size correction to the ground-state energy with respect to the magnetic flux.

\subsection{Coupled Bethe-ansatz integral equations}
We first derive the Bethe equations in the thermodynamic limit.
For convenience, we consider the half-filled system with even $N=L$ and odd $M=N/2$.

We introduce the counting functions
\begin{equation}
    \begin{aligned}
        y_L(k) &= k + \frac{1}{L}\sum_{l=1}^{M}2\arctan(\frac{\sin k-\lambda_l}{U/4}),\\
        z_L(\lambda) &= \frac{1}{L}\sum_{j=1}^{N}2\arctan(\frac{\lambda-\sin k_j}{U/4}) \\
        & - \frac{1}{L}\sum_{m=1}^{M}2\arctan(\frac{\lambda - \lambda_m}{U/2}),\label{eq:LW-yz}
    \end{aligned}
\end{equation}
with $j=1,\ldots,N$ and $l=1,\ldots, M$.
The Bethe ansatz equations \eqref{eq:LW} can then be written as
\begin{equation}
\begin{aligned}
    &Ly_L(k_j) = 2\pi I_j - \Phi,\\
    &Lz_L(\lambda_l) = 2\pi J_l.\label{eq:LW-thermo}
\end{aligned}
\end{equation}
The root densities are defined as the densities of the corresponding Bethe quantum numbers:
\begin{equation}
    \rho_L(k) = \frac{1}{2\pi}\dv{y_L(k)}{k},\quad \sigma_L(\lambda) = \frac{1}{2\pi}\dv{z_L(\lambda)}{\lambda}.\label{eq:root-density}
\end{equation}
At half filling and zero magnetization, in the thermodynamic limit,
the charge quasimomenta ${k_j}$ in the ground state are distributed over the entire Brillouin zone $(-\pi,\pi)$ with root density $\rho(k)=\lim_{L\to\infty}\rho_L(k)$.
As a consequence, unlike away from half filling, the support of $\rho(k)$ is independent of the magnetic flux $\Phi$.
The spin rapidities ${\lambda_l}$ are distributed over $(-\infty, \infty)$ with root density $\sigma(\lambda) = \lim_{L\to\infty}\sigma_L(\lambda)$.
Taking the thermodynamic limit of Eqs.~\eqref{eq:LW-yz} and \eqref{eq:LW-thermo}, we obtain the coupled Bethe-ansatz integral equations for root densities
\begin{equation}
    \begin{aligned}
        2\pi\rho(k) &= 1 + \int_{-\infty}^{\infty}d\lambda\,\frac{2\cos k\,\sigma(\lambda)(U/4)}{(\lambda - \sin k)^2+(U/4)^2},\\
        2\pi\sigma(\lambda) &= \int_{-\pi}^{\pi}dk\,\frac{2\rho(k)(U/4)}{(\lambda-\sin k)^2+(U/4)^2} \\
        &- \int_{-\infty}^{\infty}d\lambda'\,\frac{2\sigma(\lambda')(U/2)}{(\lambda-\lambda')^2+(U/2)^2}.\label{eq:root-density-limit}
    \end{aligned}
\end{equation}
The particle density and ground-state energy density are given by
\begin{equation}
    \nu = \int_{-\pi}^{\pi}dk\rho(k),\quad e_{\mathrm{gs}}= -2\int_{-\pi}^{\pi}dk\,\rho(k)\cos k.
\end{equation}

\subsection{Finite-size corrections and NLDWs}
To determine the flux dependence of the ground-state energy,
we retain the finite-size structure of the Bethe roots.
Using Eq.~\eqref{eq:LW-thermo}, the ground-state energy density can be written as
\begin{equation}
\begin{aligned}
    &e_{\mathrm{gs}}(\Phi) \\
    &= -2\int_{-\pi}^{\pi}dk\,\rho_L(k)\cos k\sum_{j=1}^{L}\delta\qty(I_j - \frac{Ly_L(k) + \Phi}{2\pi}).\label{eq:poisson-1}
    \end{aligned}
\end{equation}
For sufficiently large $L$,
the sum over the consecutive half-odd integer quantum numbers can be extended to all half-odd integers.
The Poisson summation formula then gives
\begin{equation}
    e_\mathrm{gs} = -2\int_{-\pi}^{\pi}dk\,\rho_L(k)\cos k\sum_{m\in\mathbb{Z}}e^{im[Ly_L(k) + \Phi + \pi]}.
\end{equation}
For the large-$L$ asymptotic behavior, we retain the leading finite-size corrections in Eq.~\eqref{eq:poisson-1}.
The $m=0$ term contains the finite-size dependence of the root densities,
while the leading oscillatory corrections from the Poisson summation are given by the $m=\pm 1$ terms \cite{deVegaWoynarovich}.
As will become clear from the saddle-point analysis below,
higher Fourier modes with $|m|\geq 2$ are exponentially suppressed in the large-$L$ limit and will therefore be neglected.

We first analyze the $m=0$ contribution to the NLDWs:
\begin{equation}
    D^{(2n-1)}_{0} = -2L^{2n}\int_{-\pi}^{\pi}dk\,\cos{k}\fdv[2n]{\rho_L(k)}{\Phi}\eval_{\Phi=0}.\label{eq:D-0}
\end{equation}
Here, $\rho_L(k)$ depends on the magnetic flux $\Phi$ only implicitly through the $\Phi$ dependence of the Bethe roots $k_j$.
Although this term does not contain an explicit oscillatory factor from the Poisson summation,
its flux dependence arises from the finite-size correction to the root densities.
From Eqs.~\eqref{eq:LW-yz} and \eqref{eq:root-density},
the finite-size root densities $\rho_L(k)$ and $\sigma_L(\lambda)$ satisfy
\begin{align}
    &2\pi\rho_L(k) = 1 +\cos{k}\int_{-\infty}^{\infty}d\lambda\frac{2(U/4)\sum_{l}\delta(\lambda-\lambda_l)}{(\lambda-\sin{k})^2+(U/4)^2},\notag\\
    &2\pi\sigma_L(\lambda) = \int_{-\pi}^{\pi}dk\,\frac{2(U/4)\sum_{n}\delta(k-k_n)}{(\lambda-\sin k)^2+(U/4)^2} \notag\\
        &- \int_{-\infty}^{\infty}d\lambda'\,\frac{2(U/2)\sum_{l}\delta(\lambda'-\lambda_l)}{(\lambda-\lambda')^2+(U/2)^2}.
\end{align}
We write
\begin{equation}
    \rho_L(k) = \rho(k) + r_L(k),\quad \sigma_L(\lambda) = \sigma(\lambda) + s_L(\lambda),
\end{equation}
where $r_L(k)$ and $s_L(\lambda)$ denote the leading finite-size corrections to the charge and spin root densities, respectively.
We now apply the Poisson summation formula to the distributions of the Bethe roots.
The leading finite-size corrections $r_L$ and $s_L$ are generated by the nonzero Fourier components of the root densities.
Keeping only the leading nonzero Fourier components, we obtain
\begin{align}
    &2\pi\fdv[2n]{r_L(k)}{\Phi} = \int_{-\infty}^{\infty}d\lambda\,\frac{2(U/4)\cos{k}}{(\lambda-\sin{k})^2+(U/4)^2}\fdv[2n]{s_L(\lambda)}{\Phi}\\
    &2\pi\fdv[2n]{s_L(\lambda)}{\Phi} + \int_{-\infty}^{\infty}d\lambda'\frac{2(U/2)}{(\lambda-\lambda')^2+(U/2)^2}\fdv[2n]{s_L(\lambda')}{\Phi}\notag\\
    &= \int_{-\pi}^{\pi}dk\,\frac{2(U/4)}{(\lambda-\sin{k})^2+(U/4)^2}\fdv[2n]{r_L(k)}{\Phi}\notag\\
    &+(-1)^n\int_{-\pi}^{\pi}dk\,\frac{4(U/4)\rho(k)}{(\lambda-\sin{k})^2+(U/4)^2}\cos(Ly(k)+\Phi).
\end{align}
The charge root density in the thermodynamic limit appearing in these equations is given by \cite{Essler}
\begin{equation}
    \rho(k) = \frac{1}{2\pi} + \cos{k}\int_{-\infty}^{\infty}\frac{d\omega}{2\pi}\frac{J_0(\omega)}{1+\exp(|\omega|U/2)}e^{-i\omega\sin{k}}.
\end{equation}
These equations can be solved by Fourier transforms:
\begin{align}
    \fdv[2n]{r_L(k)}{\Phi} = &4(-1)^{n}\int_{-\pi}^{\pi}\frac{dk'}{2\pi}\bigg[\rho(k')\cos(Ly(k')+\Phi)\notag\\&\times\left.\pdv{}{k}\int_{0}^{\infty}d\omega\,\frac{\sin(\omega(\sin{k}-\sin{k'}))}{\omega(1+e^{\omega U/2})}\right]\label{eq:rho-L}\\
    \fdv[2n]{s_L(\lambda)}{\Phi} = &(-1)^n\int_{-\pi}^{\pi}\frac{dk}{2\pi}\bigg[\rho(k)\cos(Ly(k)+\Phi)\notag\\
    &\qquad\qquad\times\int_{-\infty}^{\infty}d\omega\,\frac{e^{i\omega(\lambda-\sin{k})}}{\cosh(\omega U/4)}\bigg].
\end{align}
Since $\rho(k)$ is independent of the boundary condition,
the flux dependence of $\rho_L(k)$ comes entirely from $r_L(k)$.
Thus, inserting Eq.~\eqref{eq:rho-L} into Eq.~\eqref{eq:D-0}, we obtain the contribution of the $m=0$ term to the NLDWs.
Here, the integral over $k'$ is evaluated by the saddle-point approximation.
After deforming the integral contour into the complex plane,
the integral is dominated by the complex saddle point satisfying
\begin{equation}
    \dv{y_L}{k} = 2\pi\rho_L(k) = 0.
\end{equation}
In the large-$L$ limit, the saddle point is approximated by
\begin{equation}
    k_0 = \pi + i\sinh^{-1}(U/4),
\end{equation}
which satisfy $\rho(k_0) = 0$.
The saddle-point approximation yields
\begin{align}
    D_{0}^{(2n-1)} &\simeq 2(-1)^{n}L^{2n-1-1/2}\frac{\sqrt{(4/U)^2+1}e^{-L/\xi(U)}}{\sqrt{\pi|y''(k_0)|/2}} 
    \notag\\
    &\quad \times \int_{0}^{\infty}dx\,e^{-x}J_1(4x/U)\tanh{x},
\end{align}
where the correlation length is defined as
\begin{equation}
    \frac{1}{\xi(U)} = -iy(k_0) -i\pi = \frac{4}{U}\int_{1}^{\infty}dy\,\frac{\ln(y+\sqrt{y^2-1})}{\cosh(2\pi y/U)}.
\end{equation}

The $m=\pm 1$ contribution takes the form, after integration by parts,
\begin{align}
    D^{(2n-1)}_{\pm1} = 2(-1)^{n+1}\frac{L^{2n-1}}{\pi}\int_{-\pi}^{\pi}dk\,\sin k\sin(Ly_L(k)).
\end{align}
The integral is evaluated again via the saddle-point approximation, yielding
\begin{align}
    D_{\pm1}^{(2n-1)}&\simeq 2(-1)^{n+1}L^{2n-1-1/2}\frac{(U/4)e^{-L/\xi(U)}}{\sqrt{\pi|y''(k_0)|/2}}.
\end{align}
Therefore, the NLDWs have the asymptotic form
\begin{equation}
    D^{(2n-1)}(L)\sim (-1)^{n+1}D(U)L^{2n-1-1/2}e^{-L/\xi(U)},
\end{equation}
where the prefactor $D(U)$ is independent of the order of the NLDWs and given by
\begin{widetext}
\begin{align}
   D(U) 
   = \frac{2\qty(\qty(U/4)^2 - \sqrt{1+\qty(U/4)^2})\int_{0}^{\infty}dx\,e^{-x}J_1\qty(4x/U)\tanh{x}}{\qty[\frac{\pi}{2}\int_{0}^{\infty}dx\,e^{-x}J_0\qty(4x/U)\qty(\qty(U/4)^2+\qty(1+\qty(U/4)^2)x\tanh{x})]^{1/2}}.
\end{align}    
\end{widetext}

\section{Finite-flux behavior of the NLDWs\label{app:phi}}
In the main text, the NLDWs are defined by derivatives of the ground-state energy at $\Phi=0$.
One may therefore wonder whether the absence of singular behavior found at quarter filling originates from a cancellation  specific to the zero-flux point.
In this Appendix, we extend the analysis of the finite-size scaling of the NLDWs to finite values of the magnetic flux $\Phi$:
\begin{equation}
    D^{(n)}(\Phi) = L^n\pdv[n+1]{E_0(\Phi;N,M)}{\Phi}\,.
\end{equation}
\begin{figure}[h]
    \centering
    \includegraphics[width=0.8\linewidth]{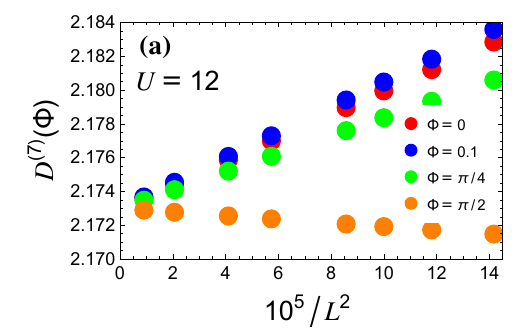}
    \hfill
    \includegraphics[width=0.8\linewidth]{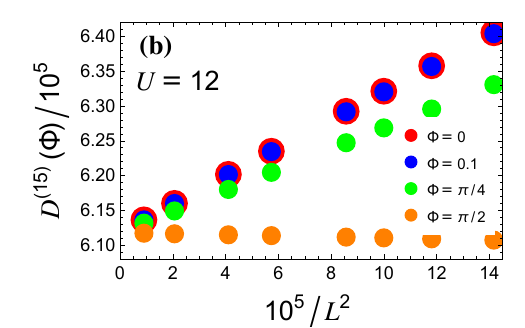}
    \caption{Finite-size scaling of (a) $D^{(7)}(\Phi)$ and (b) $D^{(15)}(\Phi)$ for $U=12$ at quarter filling.
    The NLDWs are plotted as functions of $10^5/L^2$ for $\Phi = 0, 0.1, \pi/4$, and $\pi/2$.
    For every value of the magnetic flux, the data exhibit approximately linear dependence on $L^{-2}$, indicating finite values in the thermodynamic limit.}
    \label{fig:D-phi}
\end{figure}
Figure~\ref{fig:D-phi} shows $D^{(7)}(\Phi)$ and $D^{(15)}(\Phi)$ at quarter filling as functions of $10^5/L^2$ for several values of the magnetic flux.
For all values of $\Phi$, the data exhibit approximately linear behavior and extrapolate to finite values in the thermodynamic limit.
These results indicate that the absence of divergence is not a consequence of 
an accidental cancellation at $\Phi=0$.

\section{Comparison with the XXZ chain at finite magnetization\label{app:xxz}}
We briefly discuss the spin-$\frac12$ XXZ chain in the sector with magnetization $m=-1/4$ (quarter filling in the Jordan-Wigner fermion language), 
which provides another example where the low-energy field theory predicts divergent NLDWs.

The Hamiltonian is given by
\begin{equation}
    H = \sum_{j=1}^{L}\left[\frac{J_\perp}{2}\qty(e^{i\Phi/L}S_j^+S_{j+1}^- + \mathrm{h.c.})+J_z S_j^zS_{j+1}^z\right],
\end{equation}
where $S_j^\alpha$ ($\alpha=x,y,z$) denote the spin-$1/2$ operators acting on site $j$ and $S_j^\pm = S_j^x\pm iS_j^y$.
In the following, we restrict ourselves to the gapless regime $|J_z/J_\perp|<1$.

The low-energy effective Hamiltonian is \cite{Giamarchi}
\begin{equation}
    H = H_0 + a^{\Delta-1}\lambda\int\frac{dx}{2\pi}2:\cos(8\phi):,
\end{equation}
where $H_0$ is the free-boson Hamiltonian,
\begin{equation}
    H_0 = \frac{v}{2\pi}\int_{0}^{La}dx\qty[\frac{1}{K}:\qty(\dv{\phi}{x})^2: + K:\qty(\dv{\theta}{x})^2:].
\end{equation}
Here, $K$ is the TLL parameter of the XXZ chain, $\Delta=16K$ is the scaling dimension of the Umklapp operator, $\lambda$ is normalized to have the dimension of energy, and
the bosonic fields $\phi(x)$ and $\theta(x)$ describe the low-energy degrees of freedom of the XXZ chain.

Following the same argument 
as that for the Hubbard model in Appendix~\ref{app:Umklapp},
the contribution of the Umklapp operator to the $n$th-order Drude weight behaves as
\begin{equation}
    D^{(n)}_\mathrm{Umklapp}\sim L^{n+3-2\Delta}.
\end{equation}
Therefore,
\begin{equation}
    n > 32K - 3
\end{equation}
gives the condition for the divergence of NLDWs.
\begin{figure}[ht]
    \centering
    \includegraphics[width=0.8\linewidth]{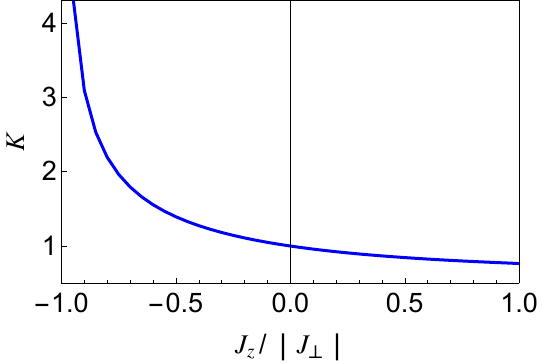}
    \caption{TLL parameter $K$ of the XXZ chain at quarter filling as a function of the anisotropy parameter $J_z/|J_\perp|$.}
    \label{fig:Kxxz}
\end{figure}
Figure~\ref{fig:Kxxz} shows the TLL parameter $K$ obtained numerically as a function of $J_z/|J_\perp|$.
Together with the condition $n>32K-3$, this figure identifies the parameter regime where field theory predicts divergent NLDWs.
\begin{figure}[h]
    \centering
    \includegraphics[width=0.8\linewidth]{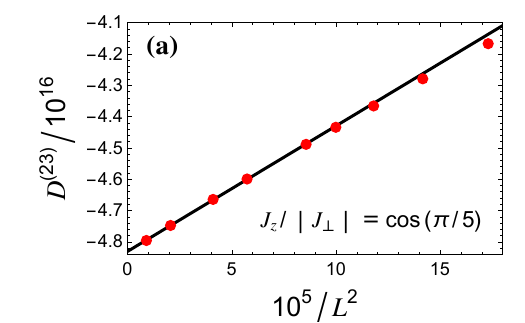}
    \hfill
    \includegraphics[width=0.8\linewidth]{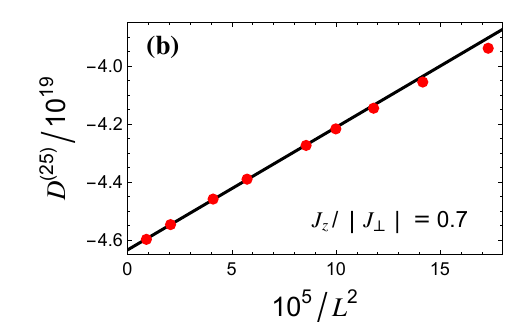}
    \caption{Finite-size scaling of the NLDWs for parameters satisfying $n>32K-3$, with $J_z/|J_\perp|$ chosen away from special points where the NLDWs at $m=0$ (half filling) remain finite for all $n$ \cite{TanikawaKatsura}. 
    We set $J_\perp=1$.
    In both cases the numerical data are linear in $L^{-2}$ for large $L$, indicating finite values in the thermodynamic limit despite the divergence predicted by field theory.}
    \label{fig:D-XXZ}
\end{figure}
Figure~\ref{fig:D-XXZ} shows the finite-size scaling of representative NLDWs for parameters satisfying $n>32K-3$.
Despite the field-theoretical prediction, the numerical data exhibit the same $L^{-2}$ finite-size scaling observed in the Hubbard model,
indicating finite values in the thermodynamic limit.
This suggests that the discrepancy between the field-theoretical prediction and the exact result is not specific to the Hubbard model but also appears in the XXZ chain.

\bibliographystyle{apsrev4-2}
\bibliography{nldw_book}

\end{document}